\documentclass[conference]{IEEEtran}
\usepackage{amsmath}
\usepackage{amssymb}
\usepackage{amsfonts}
\usepackage{amsthm}
\usepackage{algorithm}
\usepackage{algorithmic}
\usepackage{graphicx}
\usepackage{array}

\usepackage[shortlabels]{enumitem}
\usepackage{cite}
\usepackage{color}
\usepackage{epsfig,epstopdf}
\usepackage{url}
\usepackage{subfigure}
\usepackage{balance}
\usepackage{multirow}

\IEEEoverridecommandlockouts
\newcounter{MYtempeqncnt}

\begin{document}

\title{Semi-Supervised Learning with GANs for Device-Free Fingerprinting Indoor Localization}

\author{\IEEEauthorblockN{Kevin M. Chen and Ronald Y. Chang}
\IEEEauthorblockA{Research Center for Information Technology Innovation, Academia Sinica, Taipei, Taiwan}
\IEEEauthorblockA{Email: \{cwchen, rchang\}@citi.sinica.edu.tw}
\thanks{This work was supported in part by the Ministry of Science and Technology, Taiwan, under Grants MOST 106-2628-E-001-001-MY3 and MOST 109-2221-E-001-013-MY3.}}

\maketitle

\begin{abstract}
Device-free wireless indoor localization is a key enabling technology for the Internet of Things (IoT). Fingerprint-based indoor localization techniques are a commonly used solution. This paper proposes a semi-supervised, generative adversarial network (GAN)-based device-free fingerprinting indoor localization system. The proposed system uses a small amount of labeled data and a large amount of unlabeled data (i.e., semi-supervised), thus considerably reducing the expensive data labeling effort. Experimental results show that, as compared to the state-of-the-art supervised scheme, the proposed semi-supervised system achieves comparable performance with equal, sufficient amount of labeled data, and significantly superior performance with equal, highly limited amount of labeled data. Besides, the proposed semi-supervised system retains its performance over a broad range of the amount of labeled data. The interactions between the generator, discriminator, and classifier models of the proposed GAN-based system are visually examined and discussed. A mathematical description of the proposed system is also presented.
\end{abstract}

\IEEEpeerreviewmaketitle

\section{Introduction}

Many current and future Internet of Things (IoT) applications, such as smart homes, assisted living, and elderly monitoring, are enabled or facilitated by indoor location information \cite{localizationforIoT2014Macagnano,indoorsurvery2019Zafari}. To this end, fingerprint-based wireless indoor localization approaches are widely used, which involve an offline site survey phase (``fingerprinting the venue'') and an online localization phase. A reliable fingerprinting localization system that can extract and exploit the core features of the wireless signals, which are subject to environmental variations, is essential. Deep learning-based approaches to wireless indoor localization have been introduced \cite{LiuDNNjournal19, DeepFi17, DeepCNN18, HsiehCNNDNN19}. In \cite{LiuDNNjournal19}, a deep neural network (DNN) model for indoor localization was proposed and a visualization framework was developed to interpret the workings of the DNN. In \cite{DeepFi17}, a deep learning framework with a greedy learning algorithm was proposed. In \cite{DeepCNN18}, a deep convolutional neural network (DCNN) model was proposed, where the measured wireless data were transformed into the image form. In \cite{HsiehCNNDNN19}, different deep learning models and different wireless measurements for indoor localization were compared.

The aforementioned deep learning-based solutions are based on {\it supervised} learning, i.e., only labeled data collected in the site survey are used to train the fingerprinting localization system. However, data labeling is labor-intensive and time-consuming, and thus it is practically useful to utilize unlabeled data which can be collected continuously in an indoor environment with low cost. Indoor localization systems based on {\it semi-supervised} learning, which use a small amount of labeled data and a large amount of unlabeled data for training, have been proposed \cite{Pulkkinen2011semi, Gu2015semi, Ghourchian2017semi, AFDCGAN2019Li, Wang2018semi, Zhou2017semi}. In \cite{Pulkkinen2011semi}, a manifold learning technique for building accurate fingerprints from partially labeled data was proposed. In \cite{Gu2015semi}, a semi-supervised deep extreme learning machine (SDELM), exploiting semi-supervised learning, deep learning, and extreme learning machine (ELM), was developed. In \cite{Ghourchian2017semi}, a semi-supervised learning framework with two training phases was proposed. In \cite{AFDCGAN2019Li}, a generative model was used to expand the training dataset with few labeled training data. In \cite{Wang2018semi}, graph-based semi-supervised learning was proposed to reduce the data collection time for indoor localization. In \cite{Zhou2017semi}, a manifold alignment approach to reduce the overall fingerprint calibration effort was developed.

\begin{figure*}[t]
\begin{center}
\includegraphics[width=1.9\columnwidth]{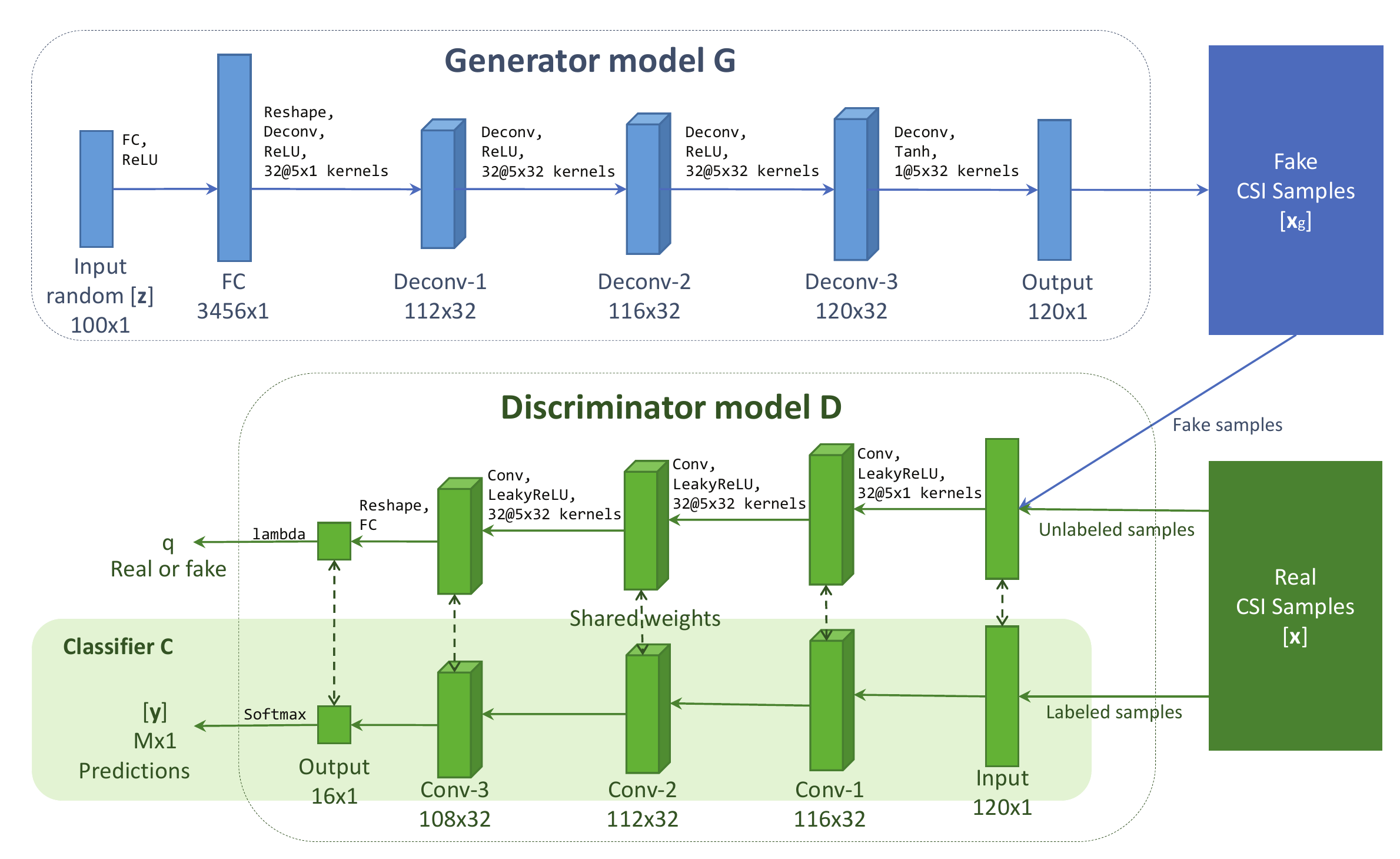}
\end{center}
\caption{Architecture overview of the proposed semi-supervised deep convolutional generative adversarial network (DCGAN) model for device-free fingerprinting indoor localization.}
\label{fig:sDCGAN_diagram}
\end{figure*}

The generative adversarial network (GAN) \cite{GAN2014Goodfellow} is a machine learning framework composed of two competing and mutually enhancing neural networks, i.e., a generator and a discriminator. A deep convolutional neural network-based architecture was introduced into GAN \cite{radford2015dcgan}, termed deep convolutional generative adversarial network (DCGAN), which greatly improves the quality of the generated artificial data. In \cite{Odena2016sGAN}, a new discriminator model that serves dual purposes was proposed, i.e., simultaneously acting as a traditional discriminator (real/fake binary classification) and a classifier (multiclass classification on a given dataset). Inspired by these studies, in this work, we propose a DCGAN-based semi-supervised device-free fingerprinting indoor localization system. In our system, the measured wireless channel state information (CSI) data are used directly, instead of being transformed into an image form as in \cite{AFDCGAN2019Li, DeepCNN18}, to avoid possible redundant dimensions (redundant image pixels) resulted from the transformation. The main contributions of this paper are summarized as follows:
\begin{itemize}
\item {\it Mathematical description:} A detailed mathematical description of the semi-supervised DCGAN model is presented.
\item {\it Practicality and performance:} The proposed semi-supervised DCGAN model achieves identical performance as the supervised state-of-the-art classifier when a large amount of labeled data is available, and achieves significantly improved performance over the supervised classifier when only a small amount of labeled data is available. The proposed model exploits the unlabeled data that can be collected continuously over time without human intervention or effort.
\item {\it Interpretation and discussion:} The interactions between the generator, discriminator, and classifier of the proposed semi-supervised DCGAN model are discussed and visually examined.
\end{itemize}

The rest of the paper is organized as follows. Section~\ref{sec:method} introduces the proposed semi-supervised learning scheme for fingerprinting indoor localization. Section~\ref{sec:results} presents experimental results and discussion. Section~\ref{sec:conclusion} concludes the paper.

\section{Semi-Supervised Learning for Device-Free Fingerprinting Indoor Localization} \label{sec:method}

We consider a 2D wireless indoor localization problem. We consider the fingerprinting approach and a device-free system (i.e., no tracking device is attached to the target to be positioned). Ambient or purpose-built wireless mediums could be used. The device-free fingerprinting indoor localization is modeled as a classification problem. The objective is to identify the location of a target, among $M$ target locations, by matching the online testing data with the offline database (``fingerprints''). Machine learning-based approaches are adopted for performing the matching. 

The architecture of the proposed semi-supervised DCGAN model for fingerprinting indoor localization is depicted in Fig.~\ref{fig:sDCGAN_diagram}. The DCGAN is composed of a Generator model G and a Discriminator model D. The training of DCGAN involves a two-step iterative process: training D and training G. In training D, the fake samples generated by G are mixed with the real samples (usually the proportion is $1:1$) as the input for D. D is trained together with a Classifier model C as a dual separate model with shared weights \cite{Salimans16}. Specifically, D is iteratively trained to perform two tasks: predict the class of the real labeled samples, and distinguish the fake samples from the real unlabeled samples. After D is trained, G is subsequently trained to generate improved fake samples. The overall training of DCGAN involves solving a minimax optimization with a properly defined loss function \cite{Salimans16}. 

The proposed model is described in detail as follows. 

\subsubsection{Generator G}

G is a deconvolutional network composed of an input layer, a fully-connected layer, $L_g=3$ deconvolutional (also called transposed-convolutional) layers, and an output layer. The input layer takes a $W_g = 100$ dimensional Gaussian random vector as its input, denoted as $\mathbf{z} = [z_1,z_2, \ldots, z_{W_g}]^{\top}$. The output of the fully-connected layer (denoted as FC) is described by
		\begin{equation}
		\mathbf{v}^{\rm (FC)} = {\rm ReLU} \left(\mathbf{T}_{}^{\rm (FC)}\mathbf{z} + \mathbf{s}^{\rm (FC)}\right)
		\label{eq:G_fc}
		\end{equation}
where 
\begin{equation}
\mathbf{T}_{}^{\rm (FC)} = \begin{bmatrix}
								t_{1,1}^{\rm (FC)}  & \cdots  & t_{1,W_g}^{\rm (FC)}   \\ 
 								\vdots   & t_{i,j}^{\rm (FC)} & \vdots            \\  
 							 t_{N_{g},1}^{\rm (FC)}  & \cdots  & t_{N_{g},W_g}^{\rm (FC)} 
							  \end{bmatrix}
\end{equation}
is the weight matrix with the $t_{i,j}^{\rm (FC)}$ element representing the weight connecting neuron $i$ in the FC layer and neuron $j$ in the input layer, $\mathbf{s}_{}^{\rm (FC)}$ is the bias vector of the fully-connected layer, ${\rm ReLU}(\cdot)$ is the nonlinear activation function, and $N_{g}^{} = 3456$. The $N_g$-dimensional output $\mathbf{v^{\rm (FC)}}$ is expressed by $H_g=32$ stacked column vectors each of size $W_g=108$ such that $\mathbf{v}^{\rm (FC)}=\big[\mathbf{v}_1^{\top}, \mathbf{v}_2^{\top},  \ldots, \mathbf{v}_{H_{g}}^{\top}\big]^{\top}$, where $\mathbf{v}_h = [{v}_{h,1},{v}_{h, 2}, \ldots, {v}_{h, W_{g}} ]^{\top}$ for $h=1,\ldots, H_g$, to fit the subsequent deconvolution operations to produce the desired dimensions of the output (fake) CSI samples.

The first deconvolutional layer (denoted as Deconv-1) works on $\mathbf{v}^{\rm (FC)}$ with $K_g^{(1)}= 32$ kernels of size $F_g^{(1)} \times D_g^{(1)}= 5 \times 32$ and stride $S=1$. The output volume of Deconv-1 is of dimensions $W_{g}^{(1)} \times H_{g}^{(1)}$, where $W_{g}^{(1)}= S(W_{g}-1)+F_{g}^{(1)} = 112$ and $H_{g}^{(1)} = K_g^{(1)} = 32$. The output in the $k$th slice (resulted from the $k$th kernel) of Deconv-1 is described as
		\begin{equation}
		\mathbf{v}_{k}^{(1)} = {\rm ReLU} \left(\mathbf{T}_{k}^{(1)}\mathbf{v}^{\rm (FC)} + \mathbf{s}_{k}^{(1)}\right)
		\label{eq:G_deconv1}
		\end{equation}
where $\mathbf{s}_{k}^{(1)}$ is the bias of the $k$th kernel of Deconv-1, and $\mathbf{T}_{k}^{(1)} = \big[\mathbf{T}_{k,1}^{{(1)}^\top}, \mathbf{T}_{k,2}^{{(1)}^\top}, \ldots, \mathbf{T}_{k,D_{g}^{(1)}}^{{(1)}^\top}\big]$ represents the transposed convolution operation with the $k$th kernel of Deconv-1, where $\mathbf{T}_{k,d}^{(1)}$ is given in \eqref{eq:G_deconv_weights} for $d=1,\ldots,D_g^{(1)}$. The $t^{(1)}_{k,d,i,j}$ element of $\mathbf{T}_{k,d}^{(1)}$ denotes the weight connecting neuron $i$ in the $d$th slice of the FC layer and neuron $j$ in the $k$th slice of the Deconv-1 layer. Note that $\mathbf{T}_{k,d}^{(1)}$ is a sparse Toeplitz matrix where all elements along a diagonal have the same value. Then, the output feature map of Deconv-1 can be described as $\mathbf{v}^{(1)}= \big[\mathbf{v}_1^{(1)^\top}, \mathbf{v}_2^{(1)^\top}, \ldots, \mathbf{v}_k^{(1)^\top}, \ldots, \mathbf{v}_{K_g^{(1)}}^{(1)^\top}\big]^\top$.

\begin{figure*}[!ht]
\setcounter{MYtempeqncnt}{\value{equation}}
\begin{align}
\mathbf{T}_{k,d}^{(1)} &= \begin{bmatrix}
t^{(1)}_{k,d,1,1} & \cdots    & t^{(1)}_{k,d,1,F_g^{(1)}} & 0                       &\cdots         & \cdots  & 0 \\ 
0         & t_{k,d,2,2}^{(1)}  & \cdots      & t_{k,d,2,1+F_g^{(1)}}^{(1)}         &  0     & \cdots        & 0 \\
\vdots    &   \vdots    &    \vdots   &     \vdots          & \vdots       & \vdots & \vdots  \\
0         &    \cdots   &      \cdots      &       0         &  t_{k,d,W_{g},W_{g}}^{(1)} & \cdots & t_{k,d,W_{g},(W_{g}-1+F_g^{(1)})}^{(1)}
\end{bmatrix} \label{eq:G_deconv_weights} \\
\mathbf{W}_{k,d}^{(1)} &= \begin{bmatrix}
w^{(1)}_{k,d,1,1} & \cdots    & w^{(1)}_{k,d,1,F_d^{(1)}} & 0 &\cdots &\cdots  & 0 \\ 
0 & w_{k,d,2,2}^{(1)}  & \cdots  & w_{k,d,2,1+F_d^{(1)}}^{(1)} & 0 & \cdots & 0 \\
\vdots&\vdots&\vdots &\vdots& \vdots & \vdots & \vdots  \\
0&\cdots&\cdots& 0 &  w_{k, d, W_{d}^{(1)}, W_{d}^{(1)}}^{(1)} & \cdots & w_{k,d,W_{d}^{(1)},(W_{d}^{(1)}-1+F_d^{(1)})}^{(1)}  
\end{bmatrix} \label{eq:D_conv_weights}
\end{align}
\hrulefill
\end{figure*}
\addtocounter{MYtempeqncnt}{2}
\setcounter{equation}{\value{MYtempeqncnt}}

Subsequent deconvolutional layer operations are performed similarly. The output volume of Deconv-$(l-1)$ is convolved with $K_{g}^{(l)}=32$ kernels of dimensions $F_{g}^{(l)} \times D_{g}^{(l)}$, where $F_{g}^{(l)}=F_{g}^{(l-1)}$ and $D_{g}^{(l)}=H_{g}^{(l-1)}$, with stride $S^{}=1$. The output volume of Deconv-$l$ is of dimensions $W_{g}^{(l)} \times H_{g}^{(l)}$, where $W_{g}^{(l)}= S(W_{g}^{(l-1)}-1)+F_{g}^{(l)}$ and $H_{g}^{(l)}=K_{g}^{(l)}$. The output in the $k$th slice (resulted from the $k$th kernel) of Deconv-$l$, for $l=2, \ldots, L_g$, is described as
		\begin{equation}
		\mathbf{v}_k^{(l)} = {\rm ReLU} \left(\mathbf{T}_{k}^{(l)}\mathbf{v}^{(l-1)} + \mathbf{s}_{k}^{(l)}\right)
		\label{eq:G_deconv2}
		\end{equation}
where $\mathbf{T}_{k}^{(l)}$ represents the transposed convolution operation with the $k$th kernel of Deconv-$l$, and $\mathbf{s}_{k}^{(l)}$ is the bias of the $k$th kernel of Deconv-$l$. The output feature map of Deconv-$l$ is given by $\mathbf{v}^{(l)}= \big[\mathbf{v}_1^{(l)^\top}, \mathbf{v}_2^{(l)^\top}, \ldots, \mathbf{v}_k^{(l)^\top}, \ldots, \mathbf{v}_{K_g^{(l)}}^{(1)^\top}\big]^\top$.

The output layer is a deconvolutional layer working with $K_g^{(L_g+1)}= 1$ kernel of size $F_g^{(L_g+1)} \times D_g^{(L_g+1)} = 5 \times 32$ with zero padding, which results in the output volume of dimensions $W_g^{(L_g+1)}\times H_g^{(L_g+1)}$, where $W_g^{(L_g+1)}= W_g^{(L_g)}= 120$ and $H_g^{(L_g+1)} = K_g^{(L_g+1)}= 1$. The deconvolution operation of the output layer is described as 
		\begin{equation}
		\mathbf{x}_{g}^{} = \tanh\left(\mathbf{T}_{}^{(L_g+1)}\mathbf{v}^{(L_g)} + \mathbf{s}^{(L_g+1)}\right)
		\label{eq:G_deconv3}
		\end{equation}
where $\mathbf{T}_{}^{(L_g+1)}$ represents the transposed convolution operation with the sole kernel in the output layer, $\mathbf{s}^{(L_g+1)}$ is the bias, and $\tanh(\cdot)$ is the hyperbolic tangent activation function. The output layer produces the fake CSI sample denoted by $\mathbf{x}_{g}^{}= \big[x_{g,1}, x_{g,2}, \ldots, x_{g,W_g^{(L_g+1)}}\big]^\top$, which has the same dimension as the real CSI samples.

\subsubsection{Discriminator D (Including Classifier C)}

D is a convolutional network composed of an input layer, $L_d=3$ convolutional layers, and an output layer. The input layer accepts the $W_{d} = W = 120$ dimensional CSI sample $\mathbf{x} = [x_1, x_2, \ldots, x_{W_d}]^{\top}$. The first convolutional layer (denoted as Conv-1) filters the input $\mathbf{x}$ with $K_{d}^{(1)} = 32$ kernels of size $F_{d}^{(1)} \times D_{d}^{(1)} = 5 \times 1$ and stride $S = 1$. The output volume of Conv-1 is of dimensions $W_{d}^{(1)} \times H_{d}^{(1)}$, where $W_{d}^{(1)}= (W_{d}-F_{d}^{(1)})/S^{}+1 = 116$ and $H_{d}^{(1)}=K_{d}^{(1)} = 32$. The output in the $k$th slice (resulted from the $k$th kernel) of Conv-1 is described as
		\begin{equation}
		\mathbf{a}_k^{(1)} = {\rm LeakyReLU} \left(\mathbf{W}_{k}^{(1)}\mathbf{x}^{} + \mathbf{b}_{k}^{(1)}\right)
		\label{eq:D_conv1}
		\end{equation}
where ${\rm LeakyReLU}(\cdot)$ is the nonlinear activation function, $\mathbf{b}_{k}^{(1)}$ is the bias of the $k$th kernel of Conv-1, and $\mathbf{W}_{k}^{(1)}=\big[\mathbf{W}_{k,1}^{(1)}, \mathbf{W}_{k,2}^{(1)}, \ldots, \mathbf{W}_{k,D_{d}^{(1)}}^{(1)}\big]$ represents the convolution operation with the $k$th kernel of Conv-1, where $\mathbf{W}_{k,d}^{(1)}$ is given in \eqref{eq:D_conv_weights} for $d=1,\ldots,D_d^{(1)}$. The $w^{(1)}_{k,d,i,j}$ element of $\mathbf{W}_{k,d}^{(1)}$ denotes the weight connecting neuron $i$ in the $k$th slice of the Conv-1 layer and neuron $j$ in the $d$th slice of the input layer. $\mathbf{W}_{k,d}^{(1)}$ is a sparse Toeplitz matrix. The output feature map of Conv-1 is described as $\mathbf{a}^{(1)}= \big[\mathbf{a}_1^{(1)^{\top}}, \mathbf{a}_2^{(1)^{\top}}, \ldots, \mathbf{a}_k^{(1)^{\top}}, \ldots, \mathbf{a}_{K_d^{(1)}}^{(1)^{\top}}\big]^{\top}$. Subsequent convolutional layer operations can be described similarly. Conv-$l$ ($l=2, \ldots, L_d$) filters the output volume of Conv-$(l-1)$, which is of dimensions $W_{d}^{(l-1)} \times H_{d}^{(l-1)}$, with $K_{d}^{(l)}=32$ kernels of dimensions $F_{d}^{(l)} \times D_{d}^{(l)}$ where $D_{d}^{(l)}=H_{d}^{(l-1)}$, and stride $S^{}=1$. The resulting output volume of Conv-$l$ is of dimensions $W_{d}^{(l)} \times H_{d}^{(l)}$, where $W_{d}^{(l)}= (W_{d}^{(l-1)}-F_{d}^{(l)})/S^{}+1$ and $H_{d}^{(l)}=K_{d}^{(l)}$. The output in the $k$th slice (resulted from the $k$th kernel) of Conv-$l$, for $l=2, \ldots, L_d$, is described as
		\begin{equation}
		\mathbf{a}_k^{(l)} = {\rm LeakyReLU} \left(\mathbf{W}_{k}^{(l)}\mathbf{a}^{(l-1)} + \mathbf{b}_{k}^{(l)}\right).
		\label{eq:D_conv2}
		\end{equation}

After the operations of all convolutional layers, the feature map produced by Conv-$L_d$ is written as $\mathbf{a}^{(L_d)}$= $[a_1^{(L_d)}, a_2^{(L_d)}, \ldots, a_{K_d}^{(L_d)}]^{\top}$, where $K_d=W_d^{(L_d)} \times H_d^{(L_d)}$. The output layer is a fully-connected layer with $N_d=M=16$ neurons whose pre-activation values ${\mathbf c}=[c_1, c_2, \ldots, c_M]^{\top}$ are computed by
		\begin{equation}
		\mathbf{c} =\mathbf{W}_{}^{(L_d+1)}\mathbf{a}^{(L_d)} + \mathbf{b}^{(L_d+1)}
		\label{eq:D_conv3}
		\end{equation}
where
\begin{equation}
\mathbf{W}_{}^{(L_d+1)} = \begin{bmatrix}
								w_{1,1}^{(L_d+1)}  & \cdots  & w_{1,K_d}^{(L_d+1)}   \\ 
 								\vdots   & w_{i,j}^{(L_d+1)} & \vdots            \\  
 							 w_{N_d,1}^{(L_d+1)}  & \cdots  & w_{N_d,K_d}^{(L_d+1)} 
							  \end{bmatrix}
\end{equation}
denotes the weights connecting neuron $i$ in the output layer and neuron $j$ in the flattened Conv-$L_d$ layer, and $\mathbf{b}^{(L_d+1)}$ denotes the bias of the output layer. Two activation functions are used for the output to serve dual purposes: discrimination and classification. For the discriminator, a customized function $\lambda\colon M\to 1$, defined by $\lambda(\mathbf{c}) = \frac{\sum_{m=1}^{M}\exp(c_m)}{\sum_{m=1}^{M}\exp(c_m)+1}$, is used to produce a scalar $q=\lambda(\mathbf{c})\in [0,1]$ which represents the probability of the input CSI sample $\mathbf{x}$ being a real sample ($1-q$ represents the probability of the input CSI sample $\mathbf{x}$ being a fake sample). For the classifier, the softmax function $\sigma\colon M\to M$, defined by $\sigma(\mathbf{c})_m = \frac{\exp(c_m)}{\sum_{i=1}^{M}\exp(c_i)}$ for $m=1,\ldots,M$, is used to produce ${\mathbf y}=\sigma(\mathbf{c})$, and the index of the largest component in ${\mathbf y}$ is the class prediction. The input layer, the $L_d$ convolutional layers, and the output layer with the customized function $\lambda$ (or softmax function $\sigma$, respectively) form the Discriminator model D (or Classifier model C, respectively), as shown in Fig.~\ref{fig:sDCGAN_diagram}.
 
\section{Results and Discussion} \label{sec:results}

	\subsection{Dataset and Models}
	
\begin{figure}[t]
	\begin{center}
	\subfigure[]{
	    \label{fig:exp_s}
	    \includegraphics[width=0.5\columnwidth]{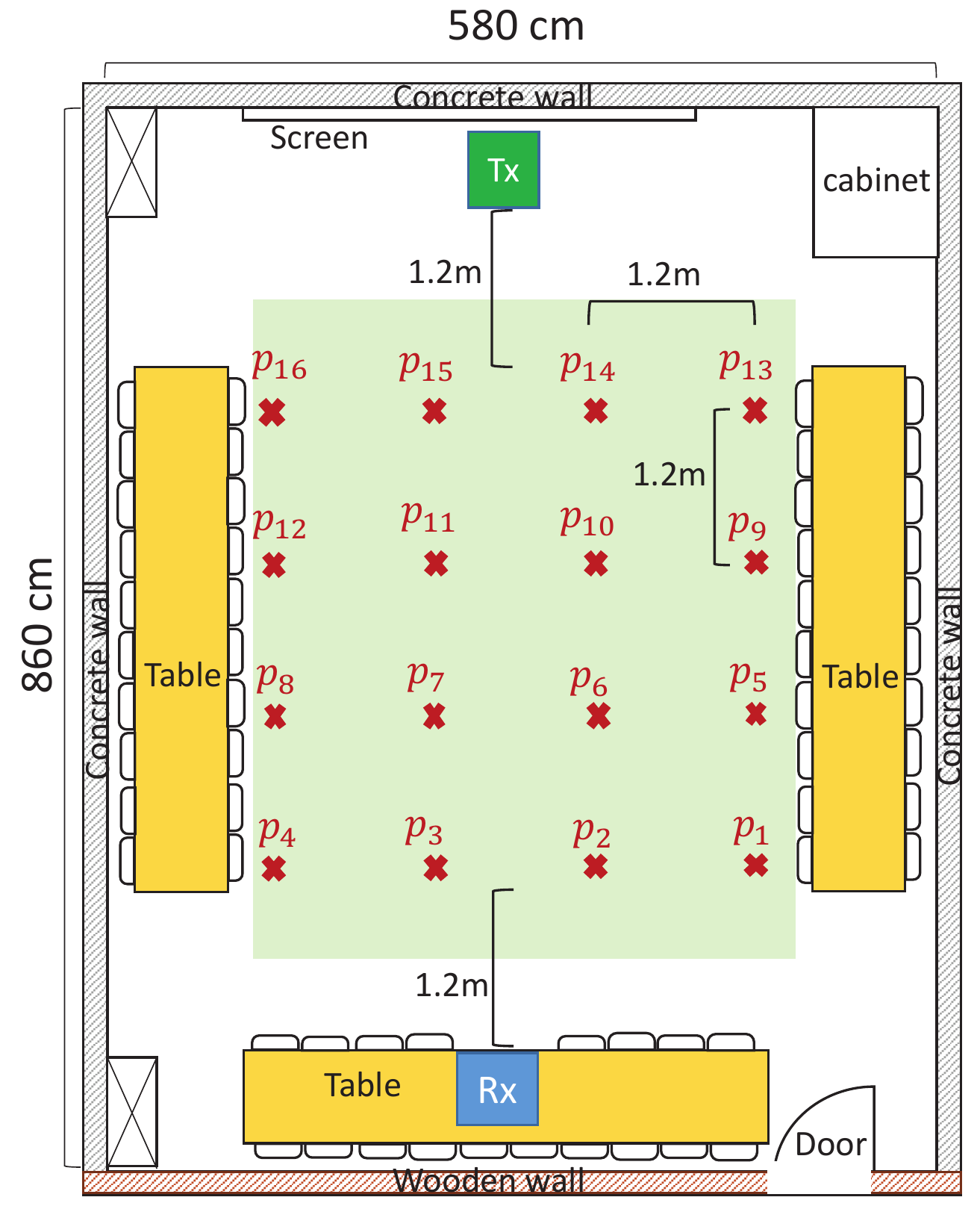}}
	\subfigure[]{
	    \label{fig:exp_p}
	    \includegraphics[width=0.453\columnwidth]{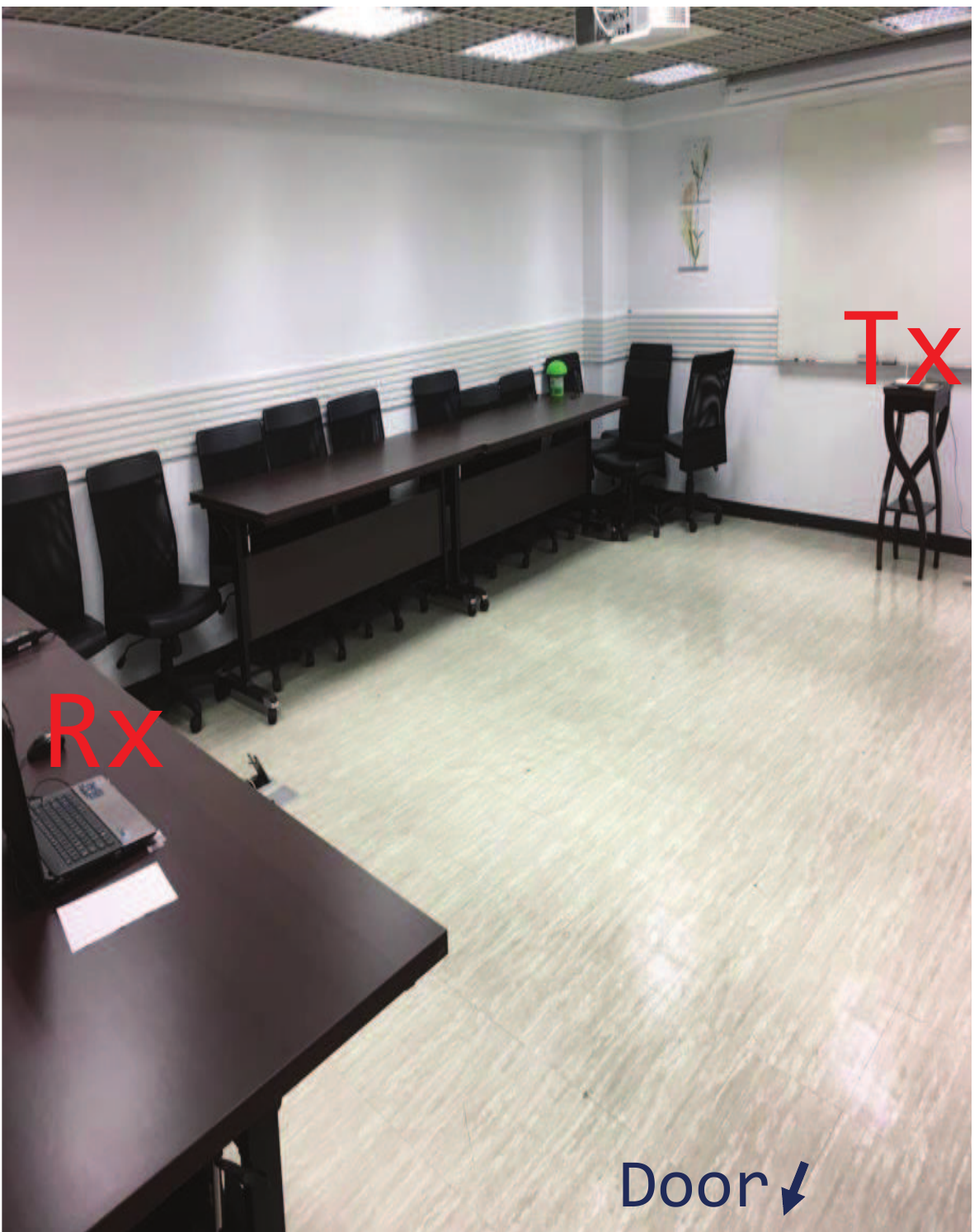}}
	\caption{(a) Floor plan and (b) photograph of the experimental scenario.}
	\label{fig:exp}
	\end{center}
	\vspace{-0.1in}
	\end{figure}

	The dataset used in this study was collected in a real indoor, conference-room-like scenario. In this scenario, a fixed-location Wi-Fi transmitter (Tx) and a fixed-location laptop receiver (Rx) were deployed in a conference room at the Research Center for Information Technology Innovation, Academia Sinica. The dimensions and layout of the environment are shown in Fig.~\ref{fig:exp}. There are $M=16$ target locations, denoted by $p_m, m=1,2,\ldots, M$. CSI samples \cite{RSSI2CSI13} were collected at the fixed-location receiver (using the tool \cite{daniel2011tool5300}) when a subject person stood at each location without any tracking device attached. The dataset contains a training set and a testing set, collected at different times and in different days. The training set has $400$ CSI samples for each location ($6400$ for all locations) and the testing set has $200$ CSI samples for each location ($3200$ for all locations). Each CSI sample is a $W=120$ dimensional vector ($30$ subcarriers with $2\times 2$ MIMO) with location label $p_m$. The CSI samples are used as {\it unlabeled} data in the model training if the label information is not used. 

	\begin{table*}[t]
	\begin{center}
	\caption{Localization Performance (in Terms of Classification Accuracy) of Semi-Supervised DCGAN and Supervised CNN}
	\label{tab:ganvscnn} \vspace*{0in}
	\begin{tabular}{|c|c|c|}
	\hline
	Labeled CSI Samples & Semi-Supervised DCGAN  & Supervised CNN   \\ \hline\hline
	16              & 85.75\% & 58.87\% \\ \hline
	32              & 85.78\% & 68.78\% \\ \hline
	64              & 87.28\% & 82.47\% \\ \hline
	128             & 87.41\% & 81.25\% \\ \hline
	1600            & 87.09\% & 86.87\% \\ \hline
	3200            & 86.72\% & 88.31\% \\ \hline
	6400            & 87.84\% & 87.71\% \\ \hline
	\end{tabular}
	\end{center}
	\vspace{-0.2in}
	\end{table*}

	\begin{table*}[t]
	\begin{center}
	\caption{Localization Performance (in Terms of Classification Accuracy) of Semi-Supervised DCGAN and Its Simplified Variant (with a Simplified G)}
	\label{tab:ganvs} \vspace*{0in}
	\begin{tabular}{|c|c|c|}
	\hline
	Labeled CSI Samples & Semi-Supervised DCGAN & Semi-Supervised DCGAN with a Simplified G \\ \hline\hline
	16              & 85.75\% & 64.40\% \\ \hline
	32              & 85.78\% & 72.94\% \\ \hline
	64              & 87.28\% & 79.25\% \\ \hline
	128             & 87.41\% & 79.41\% \\ \hline
	1600            & 87.09\% & 81.41\% \\ \hline
	3200            & 86.72\% & 86.63\% \\ \hline
	6400            & 87.84\% & 87.06\% \\ \hline
	\end{tabular}
	\end{center}
	\vspace{-0.1in}
	\end{table*}

	The training of DCGAN involves a two-step iterative process, i.e., training D/C and training G. In training D/C, first, C is trained with the full labeled training set (i.e., all $6400$ labeled real CSI samples) or reduced labeled training set. The reduced labeled training set is formed by randomly selecting an equal number of labeled real CSI samples from each location. The reduced labeled training set is of size $16, 32, 64, \ldots, 3200$ ($1, 2, 4, \ldots, 200$ labeled real CSI samples for each location). C is trained with Adam optimizer \cite{Adam15} and categorical cross-entropy loss function. Then, D is trained with the full training set with labels removed (i.e., $6400$ unlabeled real CSI samples) plus the same number of unlabeled fake CSI samples generated from G. D is trained with the Adam optimizer and binary cross-entropy loss function. After D/C is trained, G is then trained with Adam optimizer and binary cross-entropy loss function with fixed D/C, to generate improved fake CSI samples.
	
	The CNN model is adopted as the benchmark. The CNN model has the same architecture as the C in DCGAN. Also, similar to C in DCGAN, the CNN model is trained with the full or reduced labeled training set. The main difference between DCGAN and CNN is that CNN accepts labeled data only (i.e., supervised), while DCGAN can be trained with labeled data as well as unlabeled data (i.e., semi-supervised). CNN is trained with the Adam optimizer and categorical cross-entropy loss function.

	\subsection{Performance Comparison and Discussion}

	The performance of semi-supervised DCGAN and supervised CNN with equal but varying numbers of labeled real CSI samples is reported in Table~\ref{tab:ganvscnn}. When trained with sufficient labeled data (e.g., $3200$ or $6400$ labeled real CSI samples), both DCGAN and CNN achieve comparable performance, around $87\%$ accuracy. When trained with reduced amount of labeled data, CNN attains suffered performance while DCGAN retains the performance. The performance advantage of DCGAN over CNN is remarkable when as few as $16$ or $32$ labeled real CSI samples ($1$ or $2$ per location) are used. This shows the economy and robustness of DCGAN with respect to the amount of labeled data.
	
	Next, we examine the impact of the Generator G on the classification performance for DCGAN. To this end, we replace the original G in the proposed DCGAN by a simplified G comprised of only an input layer and an output layer (without the deconvolutional layers). The D/C models are intact. The training process of this simplified DCGAN is the same as the original DCGAN. Table~\ref{tab:ganvs} compares their performance with varying numbers of labeled CSI samples. It is seen that a simplified G compromises the ability of DCGAN to perform well when only limited numbers of labeled data are available. Specifically, the simplified DCGAN can no longer retain the performance when the number of labeled CSI samples reduces from $6400$ to $16$.
	
	The interaction between G and C in DCGAN is not as intuitive as that between G and D. After all, G is trained to generate improved fake samples so that D can hardly distinguish them from the real ones. However, the results here suggest that G could affect C. This may be explained as follows. When the labeled data are sufficient, C can be well trained alone (and produce good classification results) regardless of the structure of G. In contrast, when the labeled data are insufficient, C cannot be well trained alone without extra information, which is provided from G. In this case, a compromised G (in terms of architecture and trainable parameters) provides limited extra information and leads to compromised model classification performance. A sophisticated G helps train a good D, and consequently a good C, in the considered DCGAN architecture.

	\begin{figure*}[t]
	\begin{center}
	\subfigure[]{
	    \label{fig:e0[p2]}
	    \includegraphics[width=0.38\columnwidth]{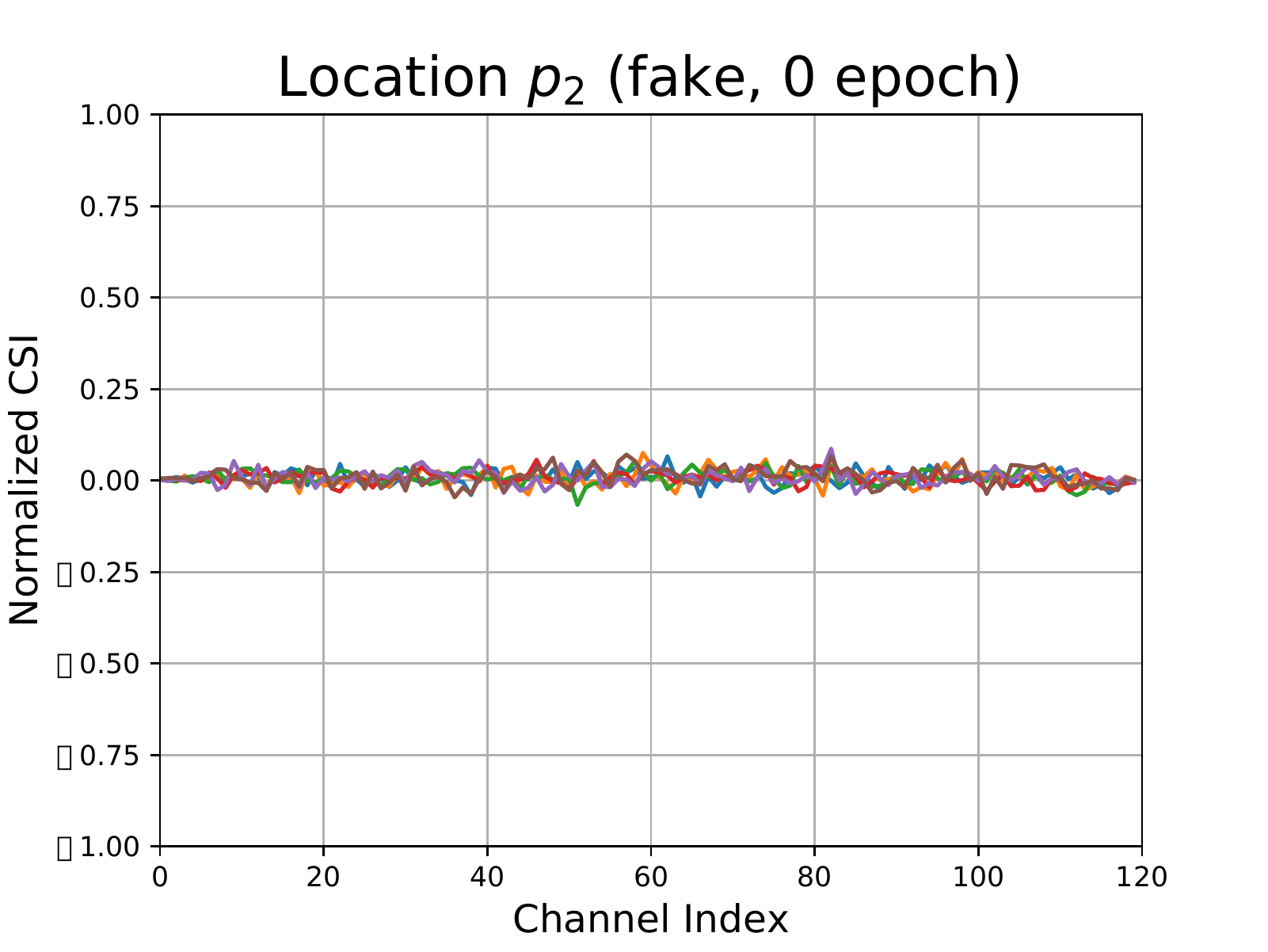}}    
	\subfigure[]{
	    \label{fig:e1[p2]}
	    \includegraphics[width=0.38\columnwidth]{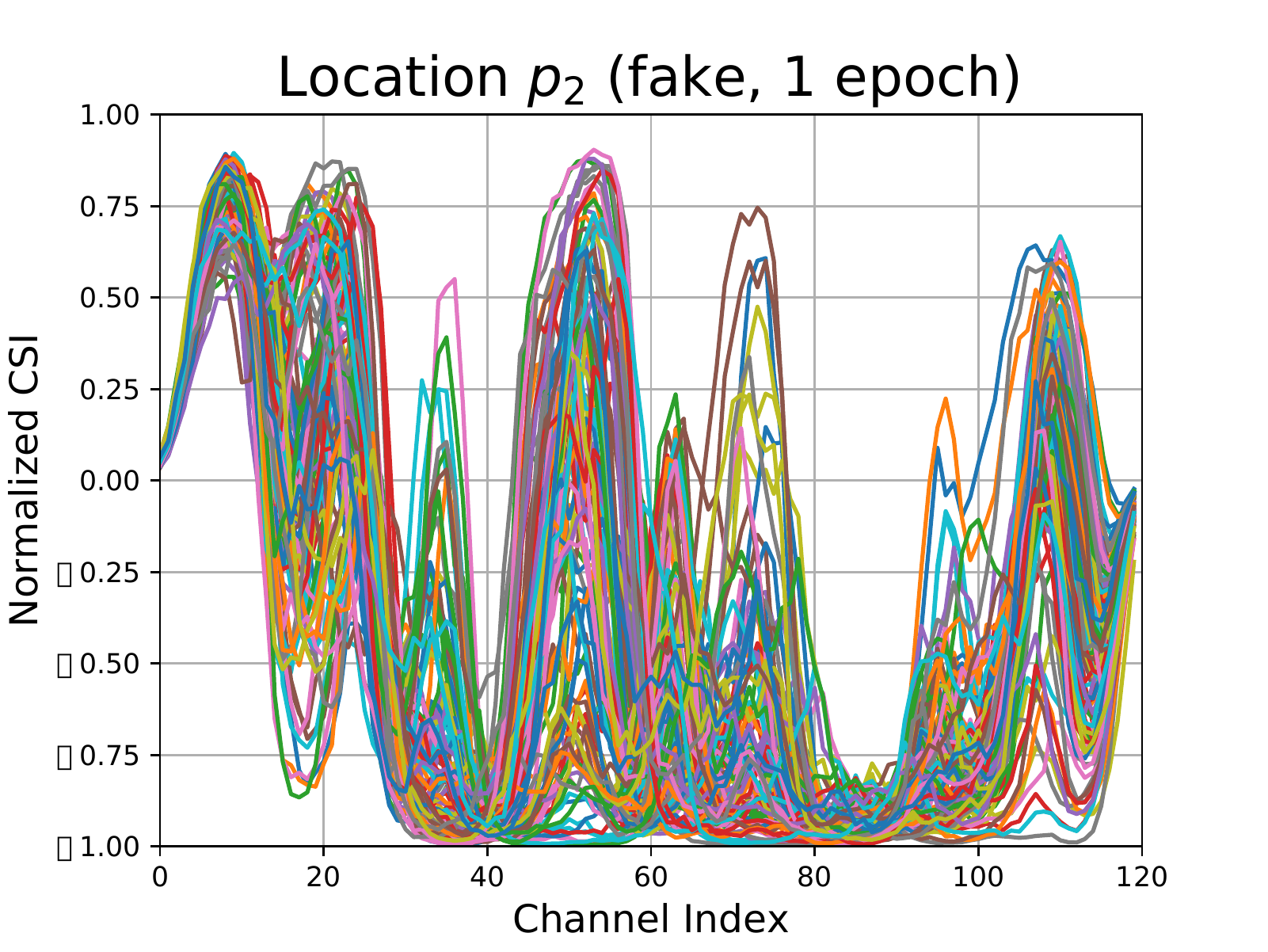}}    	    
	\subfigure[]{
	    \label{fig:e10[p2]}
	    \includegraphics[width=0.38\columnwidth]{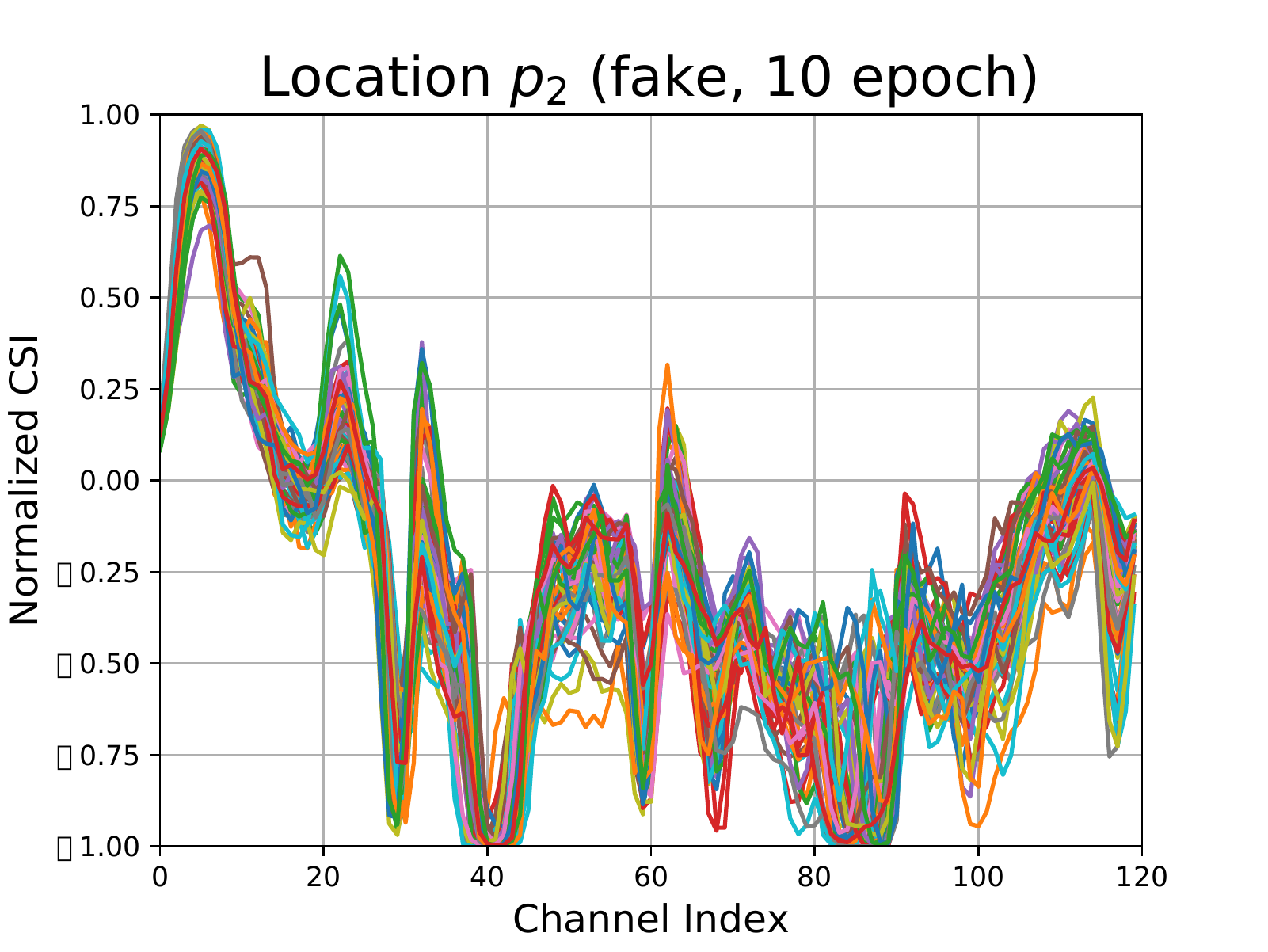}} 
	\subfigure[]{
	    \label{fig:e100[p2]}
	    \includegraphics[width=0.38\columnwidth]{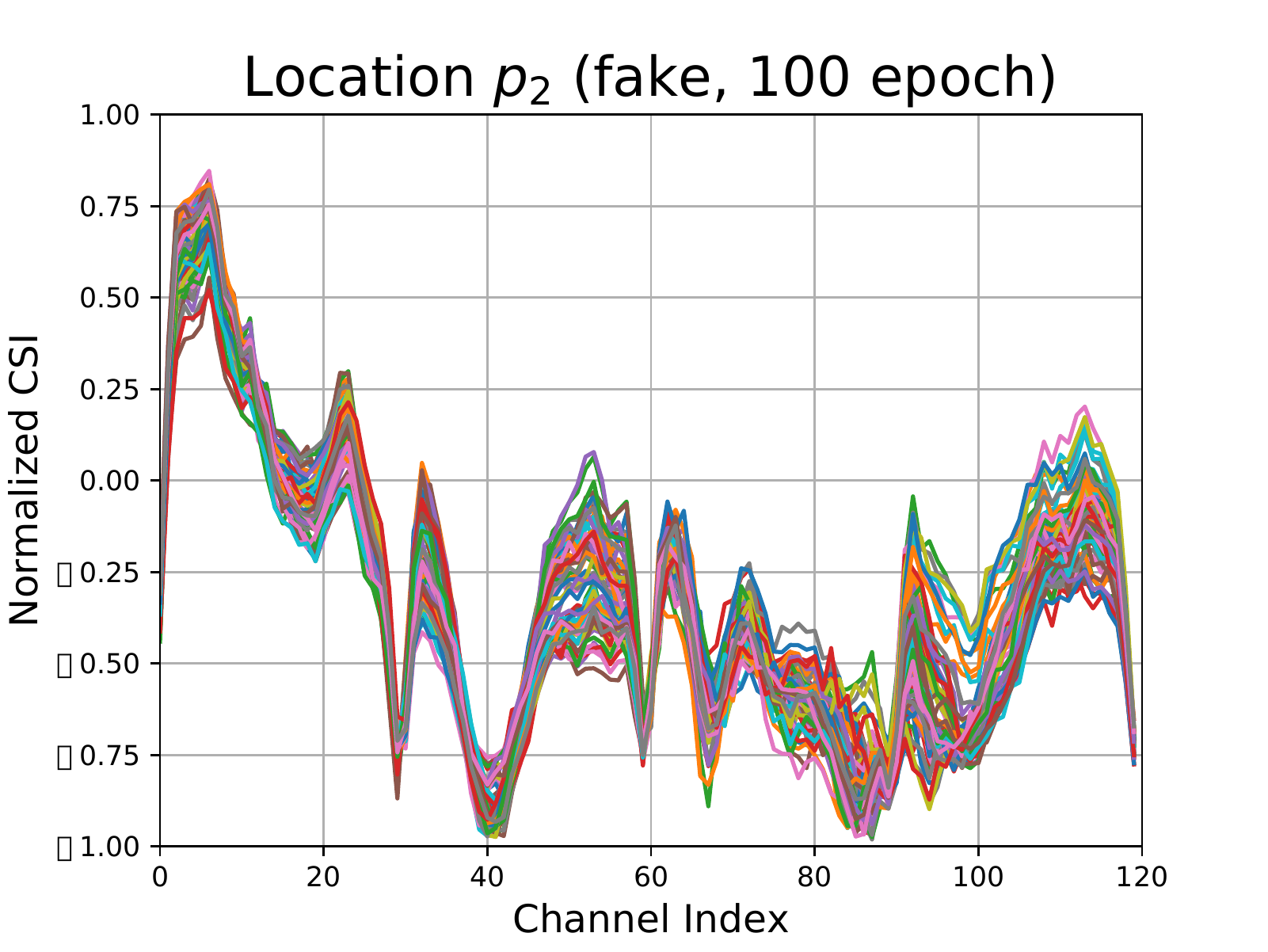}}    
	\subfigure[]{
	    \label{fig:real_p2}
	    \includegraphics[width=0.38\columnwidth]{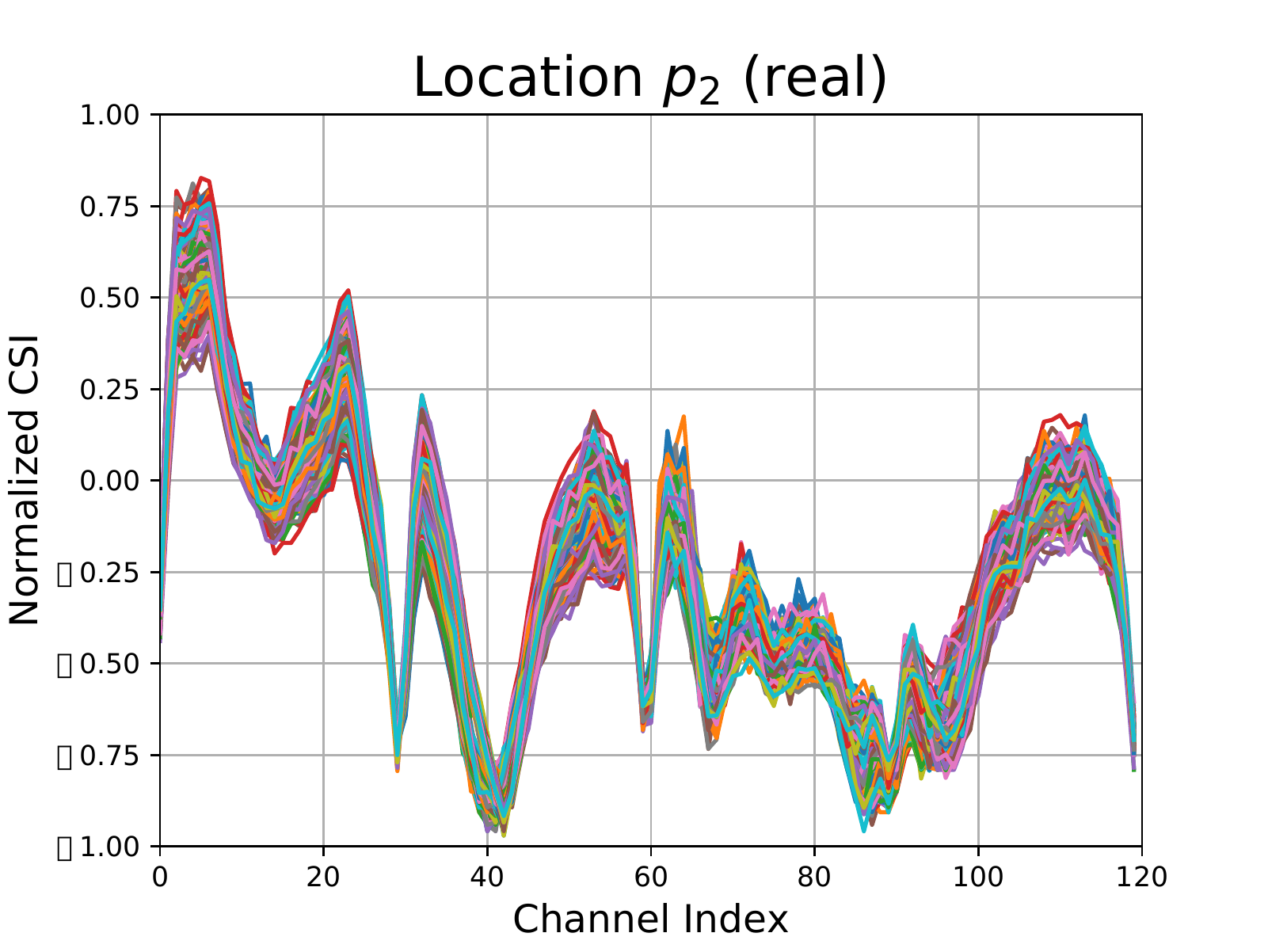}}

	\subfigure[]{
	    \label{fig:e0[p8]}
	    \includegraphics[width=0.38\columnwidth]{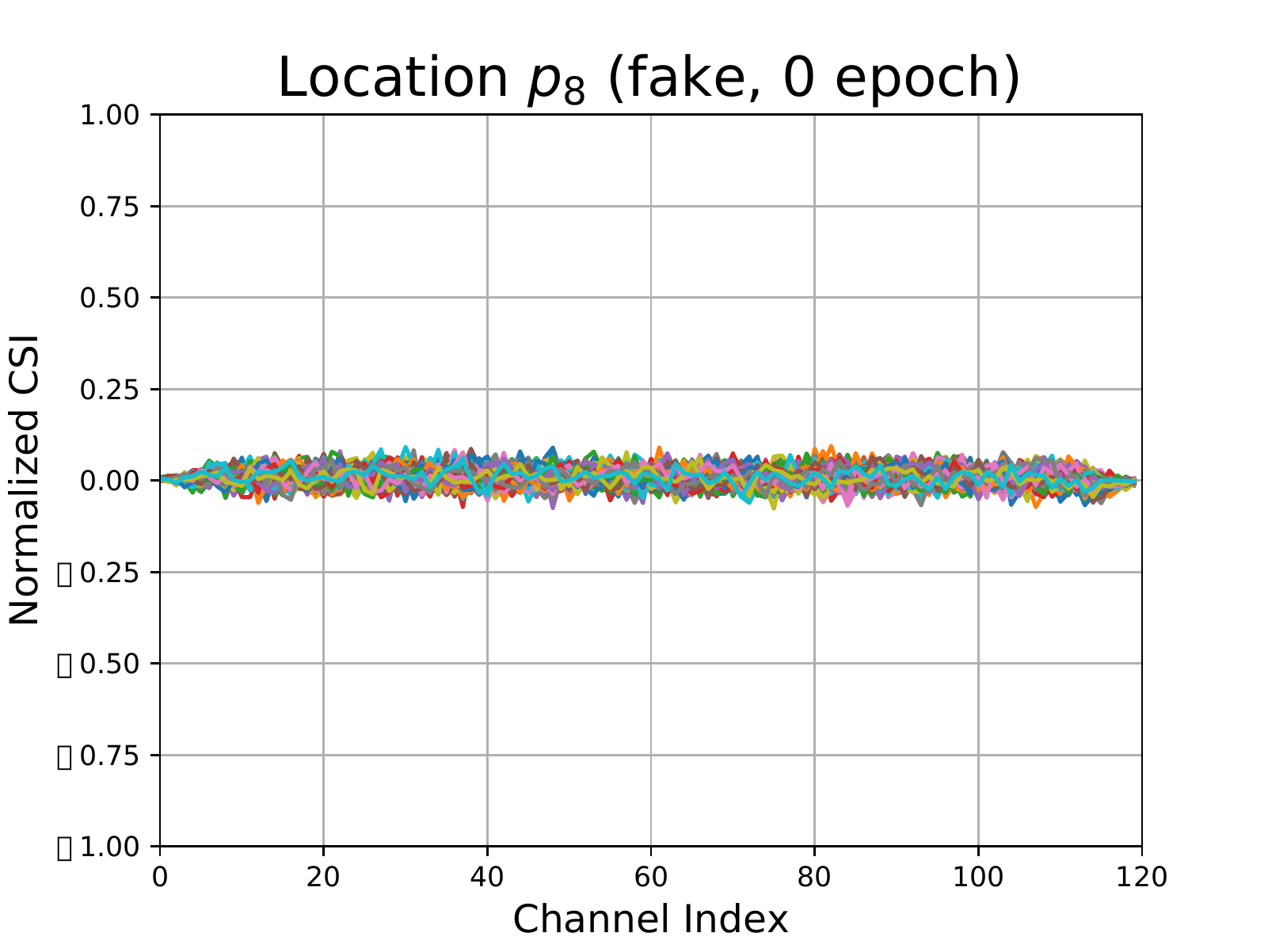}}   
	\subfigure[]{
	    \label{fig:e1[p8]}
	    \includegraphics[width=0.38\columnwidth]{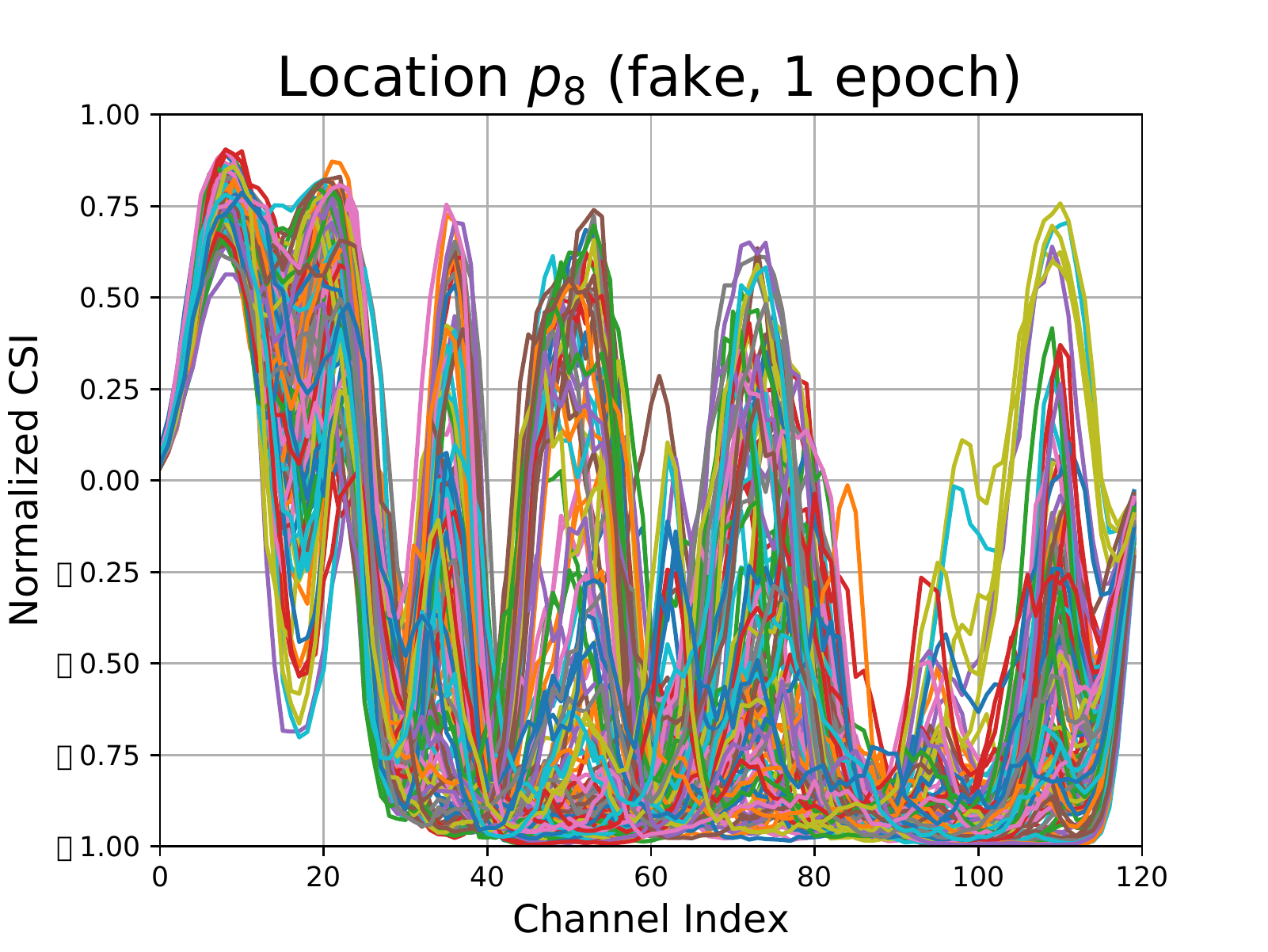}}    	 	     
	\subfigure[]{
	    \label{fig:e10[p8]}
	    \includegraphics[width=0.38\columnwidth]{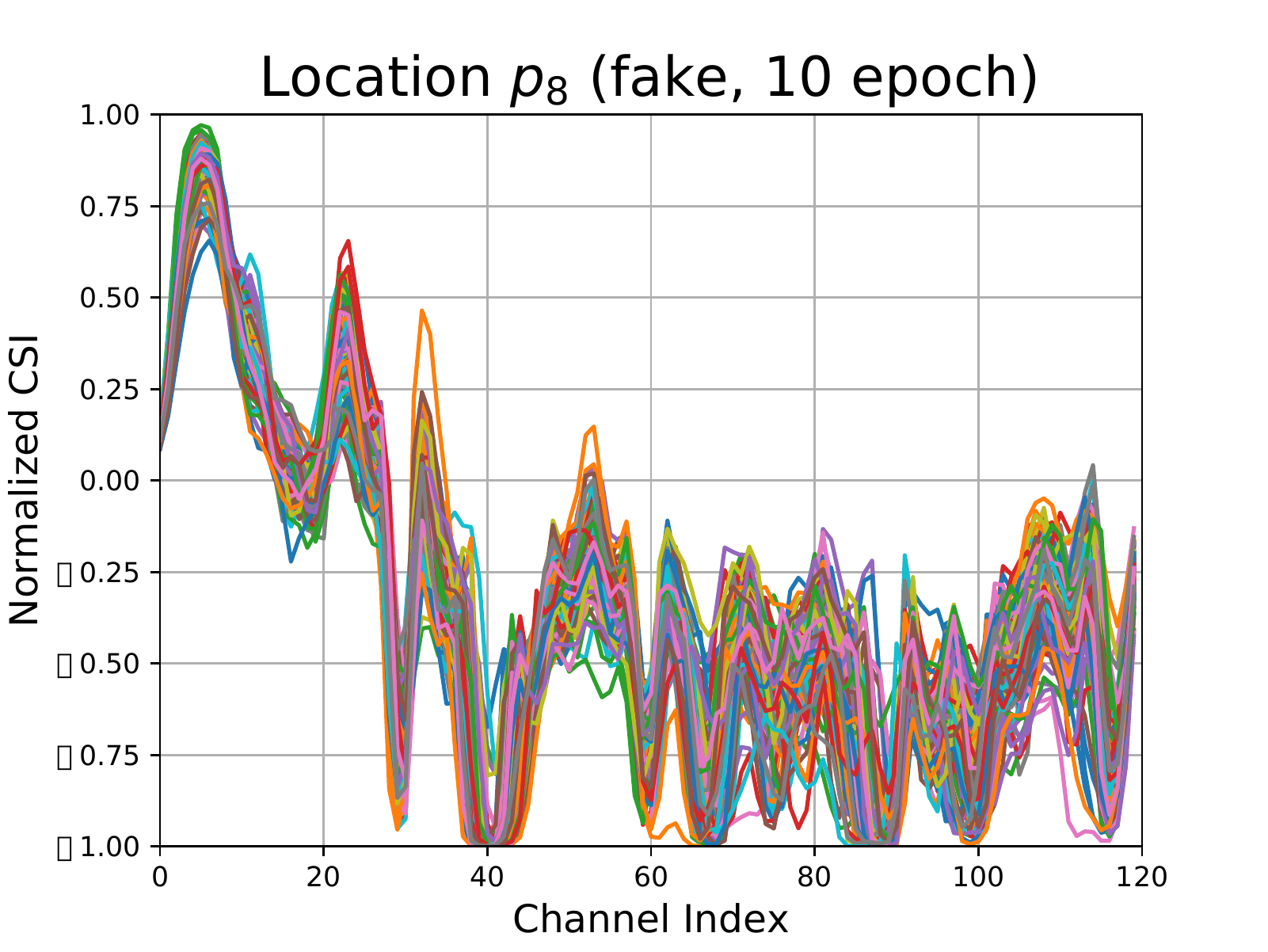}}    	    
	\subfigure[]{
	    \label{fig:e100[p8]}
	    \includegraphics[width=0.38\columnwidth]{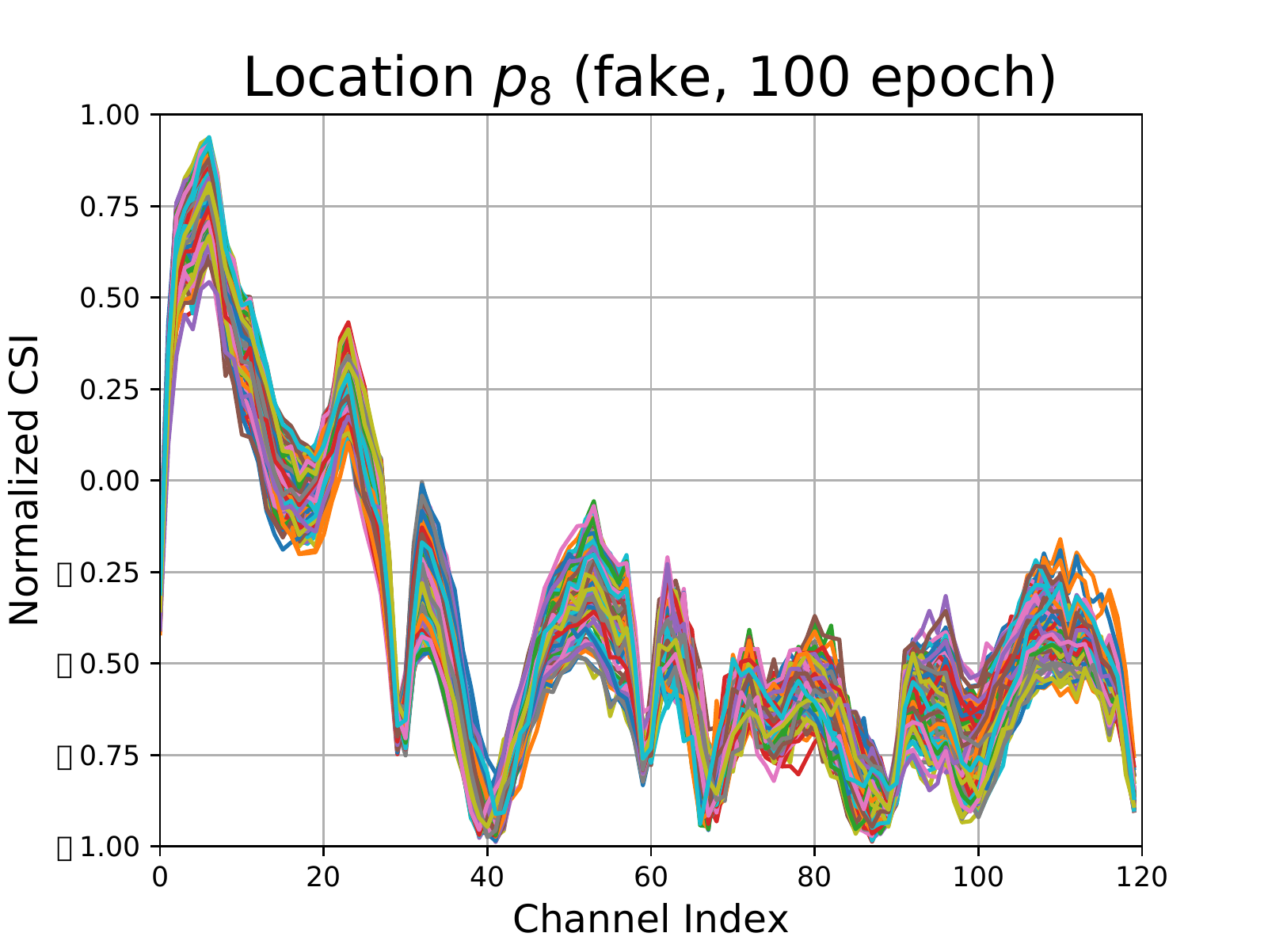}}    
	\subfigure[]{
	    \label{fig:real_p8}
	    \includegraphics[width=0.38\columnwidth]{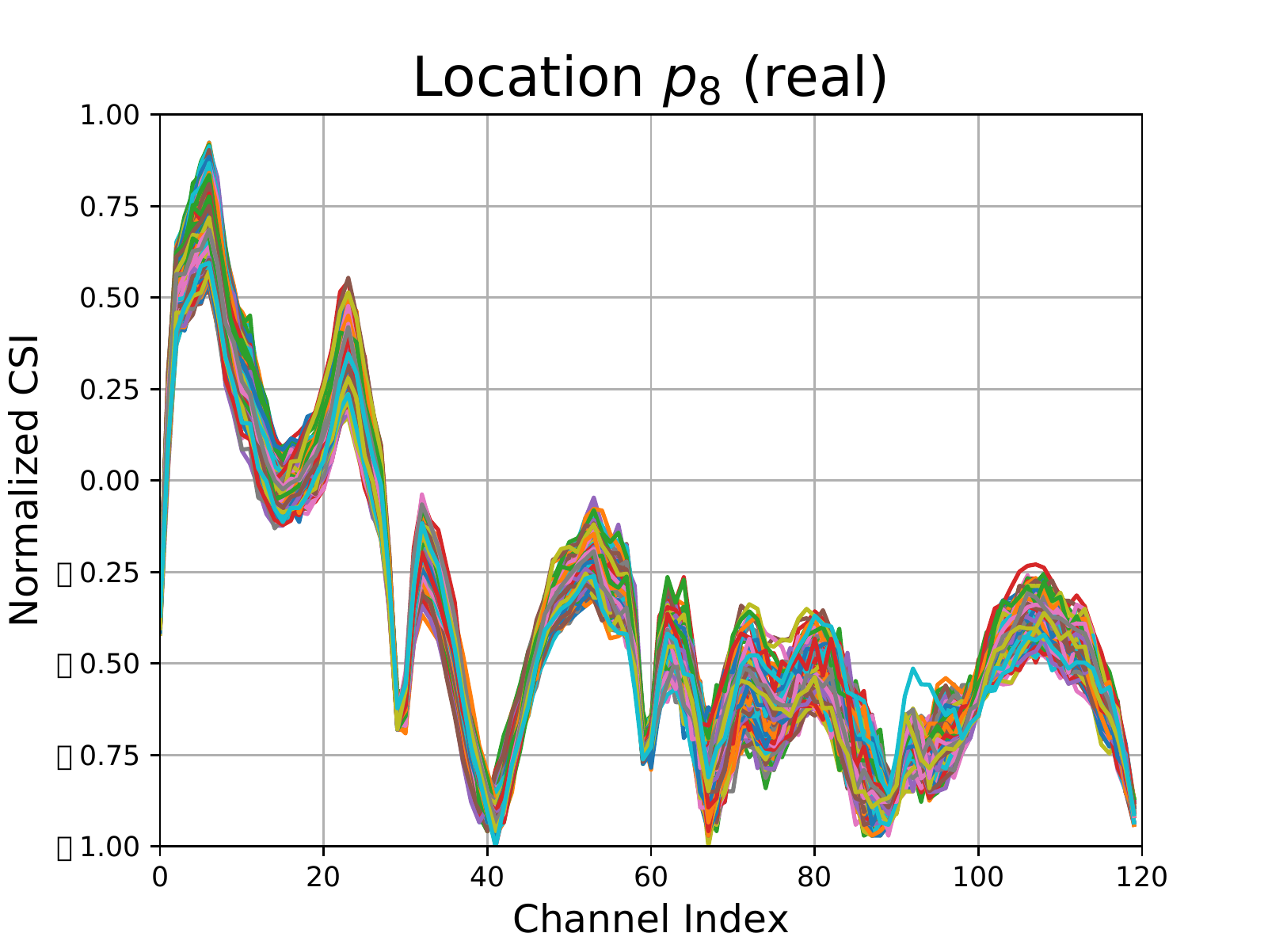}}

	\caption{(a)--(d) Fake CSI samples generated by G of DCGAN in four different epochs of training, i.e., epoch 0 (initialization), epoch 1, epoch 10, and epoch 100 (end of training), respectively, and (e) real CSI samples, for location $p_2$. (f)--(j) plot the same for location $p_8$.}
	\label{fig:realfake}
	\end{center}
	\end{figure*}

	Finally, we visually examine the progressive generation of fake CSI samples by G during training as compared to the real CSI samples. Here, the DCGAN model adopts the reduced labeled training set of size $16$. For plotting purposes, the unlabeled fake samples generated by G are assigned with the label predicted by C of the same model, and compared with the real CSI samples of the same label. The values of real and fake CSI samples are normalized to $[-1,1]$ for plotting. Fig.~\ref{fig:realfake} shows the results for two example locations: location $p_2$ (first row) and location $p_8$ (second row). For either location, Fig.~\ref{fig:realfake} plots, from left to right, the fake CSI samples generated by G in four different epochs of training, i.e., epoch 0 (initialization), epoch 1, epoch 10, and epoch 100 (end of training), and the real CSI samples, respectively. 
	
	As can be seen, at initialization, the signals are noise-like random signals. After a single training epoch, the generated fake CSI samples change drastically, but bear little resemblance to the real CSI samples. As training continues, the generated fake CSI samples appear increasingly similar to the real CSI samples. The final generated fake samples (at epoch $100$) resemble the real CSI samples fairly closely. While some distortions can be observed in between, the general patterns of the real CSI samples are replicated in the generated fake CSI samples for the same label (e.g., for location $p_2$, the peaks above zero between the $100$th--$120$th channels are seen in both real and fake CSI samples; for location $p_8$, sub-zero values for all channels after the $30$th channel are seen in both real and fake CSI samples). Besides, there are consistent (location-specific) patterns across different fake CSI samples for the same label. The results provide visual reference for the interaction between G and C in the considered DCGAN: G learns to not only generate fake CSI samples that are indistinguishable from the real ones (the objective of D), but also generate fake CSI samples that carry location-related information to aid classification (the objective of C).

\section{Conclusion} \label{sec:conclusion}

	In this paper, we have presented a GAN-based semi-supervised approach to the device-free fingerprinting indoor localization problem. We showed that the proposed scheme achieves an increasingly advantageous performance when trained with an increasingly reduced number of labeled training samples, as compared to the supervised approach. Since data labeling is costly, the results suggested a practical use case for the proposed scheme. Furthermore, the training process of the proposed model was visualized, and the interactions between the G, D, and C of the proposed model were discussed. 

\bibliographystyle{IEEEtran}
\bibliography{IEEEabrv,ref}
\balance

\end{document}